\documentclass[aps,prx,reprint,preprintnumbers,superscriptaddress,nofootinbib,longbibliography,floatfix]{revtex4-2}
\pdfoutput=1
\usepackage{rotating}
\usepackage{array}
\usepackage{amsmath}
\usepackage[normalem]{ulem}
\usepackage{slashed}
\usepackage{booktabs}
\usepackage[pdftex,table,dvipsnames]{xcolor}
\usepackage{siunitx}
\usepackage{xfrac}
\usepackage{xspace}
\usepackage{mathtools}
\usepackage{empheq}
\usepackage{multirow}
\usepackage{amssymb}
\usepackage{url}
\usepackage{comment}
\usepackage{physics}
\usepackage{color,soul}
\usepackage{bbm}
\usepackage[caption=false]{subfig}
\usepackage{adjustbox}
\usepackage{color}
\usepackage[T1]{fontenc}
\usepackage{hyperref}
\hypersetup{
  colorlinks=true,
  citecolor=blue,
  linkcolor=blue,
  urlcolor=blue
}

\definecolor{rank1}{RGB}{0,128,0}     
\definecolor{rank2}{RGB}{102,178,102} 
\definecolor{rank3}{RGB}{255,255,204} 
\definecolor{rank4}{RGB}{255,204,153} 
\definecolor{rank5}{RGB}{255,153,153} 

\newcommand{\rankcell}[2]{\cellcolor{rank#1}#2}

\newcommand{\madgraph}{\textsc{MadGraph5\_}a\textsc{mc@nlo}\xspace}
\newcommand{\baseline}{Baseline\xspace}
\newcommand{\extended}{Subjettiness\xspace}
\newcommand{\efps}{EFP\xspace}
\newcommand{\kitchenSink}{Combined\xspace}
\newcommand{\randomKitchenSink}{Random\xspace}
\newcommand{\randomEfps}{Random (EFP)\xspace}

\begin{document}

\title{Kitchen Sink Anomaly Detection}

\author{Ranit Das}
\email{das@thphys.uni-heidelberg.de
}
\affiliation{NHETC, Dept.\ of Physics and Astronomy, Rutgers University, Piscataway, NJ 08854, USA}
\affiliation{Institute for Theoretical Physics, Universit\"{a}t Heidelberg, Germany}

\author{Marie Hein}
\email{marie.hein@rwth-aachen.de}
\affiliation{Institute for Theoretical Particle Physics and Cosmology, RWTH Aachen University, Germany}

\author{Gregor Kasieczka}
\email{gregor.kasieczka@uni-hamburg.de}
\affiliation{Institut f\"{u}r Experimentalphysik, Universit\"{a}t Hamburg, 22761 Hamburg, Germany}

\author{Michael Kr\"amer}
\email{mkraemer@physik.rwth-aachen.de}
\affiliation{Institute for Theoretical Particle Physics and Cosmology, RWTH Aachen University, Germany}

\author{Lukas Lang}
\email{lukas.lang@rwth-aachen.de}
\affiliation{Institute for Theoretical Particle Physics and Cosmology, RWTH Aachen University, Germany}

\author{Radha Mastandrea}
\email{rmastand@uchicago.edu}
\affiliation{Enrico Fermi Institute, The University of Chicago, Chicago, IL 60637, USA}
\affiliation{Data Science Institute, The University of Chicago, Chicago, IL 60637, USA}

\author{Louis Moureaux}
\email{louis.moureaux@cern.ch}
\affiliation{Institut f\"{u}r Experimentalphysik, Universit\"{a}t Hamburg, 22761 Hamburg, Germany}

\author{Alexander M\"uck}
\email{mueck@physik.rwth-aachen.de}
\affiliation{Institute for Theoretical Particle Physics and Cosmology, RWTH Aachen University, Germany}

\author{David Shih}
\email{shih@physics.rutgers.edu}
\affiliation{NHETC, Dept.\ of Physics and Astronomy, Rutgers University, Piscataway, NJ 08854, USA}

\begin{abstract}

An enormous amount of R\&D effort has resulted in many new resonant anomaly detection methods being proposed in recent years. However, the vast majority of previous R\&D studies have suffered from two limitations: they have focused 
on a very small set of simulated signal benchmark models; and they have either used small sets of carefully crafted high-level jet substructure observables, which can be highly performant but are prone to model dependence, or the full collider event phase space, which is more agnostic but suffers from reduced sensitivity. 
In this work, we address both limitations: we formulate a number of new simulated signal benchmarks, which we make publicly available in a format fully compatible with the LHCO R\&D benchmark; and we  explore a high-level, yet highly agnostic, observable set consisting of Energy Flow Polynomials in addition to the usual subjettiness variables.
We evaluate this ``kitchen sink'' observable set for both an idealized anomaly detector and the CWoLa hunting task, along with three baseline observable sets (the Baseline LHC Olympics set, subjettiness observables, and Energy Flow Polynomials). We find that our kitchen sink approach is the most sensitive to a broad range of signal types.
Furthermore, we show that an attribute bagging variant, in which each ensemble member is trained on a random subset of substructure observables, yields comparable anomaly detection performance while significantly reducing training cost.

\end{abstract}

\maketitle

\section{Introduction}

In recent years, a new paradigm of model-agnostic anomaly detection has emerged at the LHC~\cite{Kasieczka:2021xcg, Aarrestad:2021oeb,karagiorgiMachineLearningSearch2022,Belis:2023mqs,hepmllivingreview}, which aims to provide much wider coverage of Beyond the Standard Model (BSM) physics compared to the traditional toolkit of 
model-specific new physics searches. Among these approaches, resonant anomaly detection with weakly supervised methods has shown particular promise, as it has a built-in guarantee of asymptotic optimality and can be sensitive to small signal cross sections.
Therefore, weakly supervised anomaly detection has become increasingly popular over recent years, including an application in several experimental analyses~\cite{ATLAS:2020iwa,CMS:2024nsz,ATLAS:2025obc,Gambhir:2025afb,CMS:2025sch}.

A key challenge in weakly supervised approaches is that their degree of model independence is not intrinsic, but instead strongly tied to the choice of input features. In jet-based searches, potential new physics signatures are encoded in the substructure of hadronic jets. While low-level inputs such as particle four-momenta are, in principle, maximally model agnostic, they are often outperformed in practice by carefully chosen high-level observables~\cite{Buhmann:2023acn, Sengupta:2023vtm, Mikuni:2024qsr}, such as subjettiness~\cite{Thaler:2010tr, Thaler:2011gf} variables.

This tension motivates the exploration of richer and more systematically constructed high-level feature sets. In this work, we take a significant step in this direction by using Energy Flow Polynomials (EFPs)~\cite{Komiske:2017aww} as anomaly detection features. EFPs form a complete, systematically improvable basis for infrared- and collinear-safe jet observables, and therefore provide a particularly well-motivated and highly expressive description of jet substructure. To our knowledge, this is the first study to employ EFPs in a weakly supervised anomaly detection setting, thereby substantially extending the space of physically motivated, model-agnostic input features.

In order to more comprehensively and model-agnostically benchmark the performance of EFPs and other feature sets, we introduce a new suite of BSM signals for use in benchmarking resonant anomaly detection methods in this work.
Inspired by the latest model agnostic CMS search~\cite{CMS:2024nsz}, these BSM signals represent a
large and diverse set of dijet resonances with different hadronic decay topologies, resulting in substantially varied jet substructure patterns. 
The signals are designed to be fully compatible with the LHC Olympics 2020 R\&D data set~\cite{LHCOdataset} background and are published in the same formats, allowing for easy use with existing pipelines. 
We publicly release these signal benchmarks on Zenodo~\cite{signal-zenodo} for greater use by the anomaly detection community.

Within this expanded setting, we investigate a ``kitchen sink'' strategy that combines multiple classes of high-level observables -- including EFPs and subjettiness variables -- into a single, large feature set. This allows us to further limit the assumptions we inject into an anomaly detection analysis by placing the decision of which features are relevant into the hands of the anomaly detector. 
In the statistics literature, ``kitchen sink'' models are often viewed critically: while they reduce the risk of omitting relevant features, they can also introduce less informative inputs, complicating interpretation and increase overfitting if not properly regularized. 
In our case, however, the primary goal is discovery sensitivity and model coverage rather then interpretability.
Any resulting look-elsewhere effects incurred because of the larger feature space can be eliminated by the use of an independent test set~\cite{Hein:2025ysv}.
We find that this approach consistently provides the most robust
performance, yielding on average a factor of $\sim 2.5$ in increased signal sensitivity for a $5\sigma$ discovery compared to the LHC Olympics Baseline feature set across the wide range of signal models considered.
This supports the idea that maximizing the coverage of physically motivated observables is key to achieving practical model agnosticism.

Handling such large feature sets presents both conceptual and technical challenges. Recent work has demonstrated that Boosted Decision Trees (BDTs) are well suited for high-dimensional inputs composed of many high-level observables~\cite{Finke:2023ltw, Freytsis:2023cjr,grinsztajn2022treebased}. Building on this insight, we show that even very large collections of EFPs -- comprising up to $\mathcal{O}(10^3)$ features -- can be effectively exploited by BDTs. Unlike unstructured or noise-like feature expansions, these physically grounded observables lead to tangible performance gains, improving the sensitivity to small signal injections in resonant anomaly detection beyond previous benchmarks~\cite{Finke:2023ltw}.

While this feature set is highly performant, the training time of BDTs increases significantly with the feature set size. 
To enable the use of ever-larger feature sets for analysis, we therefore also propose to use a random-subsampling approach to the kitchen sink strategy. This approach still provides a robust and highly sensitive performance while significantly reducing computation time. 

We study this strategy in both idealized and realistic scenarios, considering anomaly detection with a perfect background template (BT) as well as in the CWoLa hunting framework~\cite{Metodiev:2017vrx,Collins:2018epr,Collins:2019jip}, where sideband data is used to construct the BT. In contrast to more elaborate background modeling techniques based on generative machine learning~\cite{Hallin:2021wme,Raine:2022hht,Hallin:2022eoq,Golling:2022nkl,Golling:2023yjq}, CWoLa hunting can be easily used with any number of input features and provides sufficiently accurate background estimates even in the high-dimensional feature spaces considered here. We evaluate the performance of various resonant anomaly detection methods and feature sets on a significantly expanded suite of BSM signals, providing an extensive exploration of signal diversity in this context. 

This paper is structured as follows. In Section~\ref{sec:Methods}, we introduce weakly supervised anomaly detection and describe our analysis setup. Section~\ref{sec:Data} presents the datasets used in this study, including the extended set of signal models. In Section~\ref{sec:Feature_sets}, we detail the input feature constructions. The performance of the different feature sets across the various signal models is presented in Section~\ref{sec:Results}. We conclude in Section~\ref{sec:Conclusion}, with additional studies of reduced feature sets provided in the Appendix.

\section{Methods}
\label{sec:Methods}

\subsection{Weakly supervised anomaly detection}

According to the Neyman-Pearson lemma~\cite{Neyman:1933wgr}, the optimal model-agnostic anomaly score for a given set of descriptive features $\vec{x}$ is given by the likelihood ratio between the data and background ($B$), 
\begin{equation}
    R_{\text{optimal}}(\vec{x}) = \frac{p_{\text{data}}(\vec{x})}{p_{B}(\vec{x})}.
    \label{eq:OptimalAnomalyScore}
\end{equation}

In practice, $p_\text{data}$ and $p_\text{B}$ are unknown. Hence, the optimal anomaly score can be approximated using a binary classifier trained to distinguish data in a given signal region (SR) from events in a background template (BT), which should follow the true background distribution in the SR as closely as possible. 

In this work, we focus on resonant anomaly detection, where the SR is defined as a window on a resonant variable such as the dijet invariant mass $m_{JJ}$. 
In this context, a variety of methods~\cite{Metodiev:2017vrx,Collins:2018epr,Collins:2019jip,Nachman:2020lpy,Andreassen:2020nkr,1815227,Hallin:2021wme,Raine:2022hht,Hallin:2022eoq,Golling:2022nkl,Golling:2023yjq, Das:2024fwo, Leigh:2024chm, Oleksiyuk:2025pmu}
have been developed to construct the BT from simulation or in a data-driven way.
Since we focus on feature selection and a variety of signal models in this work, we use the simplest BTs:
The idealized anomaly detector (IAD) introduced in Section \ref{sec:IAD} as an optimal benchmark and the CWoLa hunting approach introduced in Section \ref{sec:CWoLaHunting}.

In a realistic analysis, the position of the resonance in $m_{JJ}$ would not be known a priori, and one would need to scan the SR over $m_{JJ}$. 
For simplicity, in this work we center the SR at the resonance mass to compare the feature sets for various signal models in an optimal setting.

\subsubsection{The Idealized Anomaly Detector}
\label{sec:IAD}
The idealized anomaly detector (IAD) \cite{Hallin:2021wme} uses a perfect background template constructed by sampling from the true background distribution in the SR. Of course, such an ideal template is not available in real data, but it can be generated for simulated data. Therefore, the IAD simply provides an optimal benchmark for weakly supervised anomaly detection. 

\subsubsection{CWoLa hunting}
\label{sec:CWoLaHunting}

CWoLa hunting~\cite{Metodiev:2017vrx,Collins:2018epr,Collins:2019jip} is a fully data-driven weakly supervised anomaly detection method that uses the data in the sidebands (SBs) of the SR as the BT. The sidebands in our specific analysis are defined in Section \ref{sec:TrainingTesting}.

CWoLa hunting relies on two crucial assumptions. The first is the standard bump hunt assumption that the new physics resonance should be localized to the SR. The second is that the feature space $\vec{x}$ should be uncorrelated with the resonance variable $m_{JJ}$ used to define the SR and the SBs. Typically, mildly correlated observables do not significantly impact the performance of CWoLa hunting. However, more sophisticated methods for generating the BT are able to handle stronger correlations. 

\subsection{Classifier}
\label{sec:Classifier}
We use the exact same classifier setup as in Ref.~\cite{Finke:2023ltw}, the \texttt{HistGradientBoostingClassifier} from the \texttt{scikit-learn}~\cite{scikit-learn} library, which is based on the LightGBM~\cite{keLightGBMHighlyEfficient2017} implementation of Gradient Boosted Decision Trees (GBDT). 
GBDT models are a natural choice for model-agnostic anomaly detection using high-level features since they have been established to perform well on tabulated data~\cite{Finke:2023ltw,Hein:2025uhj}. We largely use the default hyperparameters of the \texttt{scikit-learn} implementation, but set the maximum number of iterations to $200$ and use a training-validation split of $50-50$. 
To stabilize and improve performance, each classifier is an ensemble over $50$ independent GBDT models by averaging their predictions.

\subsection{Evaluation metrics}
\label{sec:Metrics}

A good anomaly detector is supposed to allow for a  $5\sigma$ discovery even if the number of signal events $S$ in the signal region among $B$ background events has a much smaller starting statistical significance $\sigma_{0}=S/\sqrt{B}$. Hence,
in this work, the main performance metric for anomaly detection is the minimum initial signal significance $\sigma_{min}=S_{min}/\sqrt{B}$ at which a discovery is still possible. 
In order to approximate this expected gain in sensitivity, the significance improvement characteristic (SIC)
\begin{equation}
    \text{SIC} = \frac{\epsilon_S}{\sqrt{\epsilon_B}}\, ,
    \label{eq:SIC}
\end{equation}
is typically used, where $\epsilon_{S/B}$ denote the signal and background efficiency, respectively, i.e.\ the fraction of events passing a particular cut on the classifier score.
This approximation is a common performance measure for anomaly detection, which becomes exact in the Gaussian limit where $\epsilon_S S \ll \epsilon_B B$. 
The SIC depends on the working point (choice of threshold on the anomaly score), and for concreteness we will consider the ideal threshold that maximizes SIC\footnote{Since the SIC is dominated by statistical fluctuations at small signal and background efficiencies, we only take into account classifier thresholds for which the relative statistical error of the background efficiency is smaller than $20\%$.} and denote it by max(SIC).
One could also consider other working points e.g.\ thresholds at fixed background or data efficiencies, but that does not qualitatively change the results.

The SIC of course depends on the initial signal significance $\sigma_0$. If $\sigma_0$ is too small the classifier can no longer detect the difference between data and background, and max(SIC) drops to 1. 
We investigate this behavior in Sections \ref{sec:ResultsIAD}, \ref{sec:IADBeyondLHCO}, and \ref{sec:ResultsCWoLa}.

However, we do not use the max(SIC) to obtain the minimum initial significance $\sigma_{ min}$, because the Gaussian approximation usually overestimates the actual significance after the classifier cut, since $\epsilon_S S \ll \epsilon_B B$ does not hold.
To obtain a statistically reliable estimate of $\sigma_{min}$, we compute the Asimov estimate~\cite{Cowan:2010js} 
\begin{equation}
    \mathcal{S} = \sqrt{2 [(s + b) \ln\left(1+\frac{s}{b}\right)-s]} \, ,
    \label{eq:AsimovStatistic}
\end{equation}
where $s=S \cdot \epsilon_S$ and $b=B\cdot \epsilon_B$ are the number of signal and background events in the SR after the given classifier cut, respectively.
As for the SIC, we use the maximum of the Asimov estimate over all working points.
Hence, we determine $\sigma_{min}$ by solving the equation\footnote{
In practice, we obtain this function by training $n$ different classifier with different initial significances $\sigma_{0,0}<\sigma_{0,i}< \sigma_{0,n}$, to determine $\text{max}(\mathcal{S}(\sigma_{0,i}))$ and interpolate piecewise linearly to construct a continues function $\max(\mathcal{S})(\sigma_0)$.
} 
\begin{equation}
    \text{max}(\mathcal{S})(\sigma_{min}) = \sigma_t\, ,
    \label{eq:minIS}
\end{equation}
where $\sigma_t$ is the target significance, typically at $3\sigma$ or $5\sigma$. 
In contrast to a complete analysis of all statistical and systematic effects as done in~\cite{CMS:2024nsz}, we only perform a statistical analysis in order to derive limits for different feature sets.

We construct several metrics to capture model agnosticism for the different feature sets considered. We use the average and maximum of $\sigma_{min}$ for all investigated signal models. 
Furthermore, we define the regret for a given feature set $f$ as 
\begin{equation}
    r_f = \frac{\sigma_{min,f}}{\sigma_{min,best}}\, ,
    \label{eq:regret}
\end{equation}
where $\sigma_{min,\ best}$ is the lowest minimal initial significance of the investigated feature sets.
Hence, the regret is a metric capturing how much worse a given feature set performs on a signal compared to the best tested feature set.
Model agnosticism is again estimated by taking the average and maximum of the regret for all investigated signal models.

\section{Data}
\label{sec:Data}

\subsection{LHC Olympics R\&D data set}

We use the LHC Olympics 2020 (LHCO) R\&D \cite{Kasieczka:2021xcg, LHCOdataset} data set, which is a standard benchmark in the anomaly detection literature.
This data set was produced using \textsc{Pythia 8}~\cite{10.21468/SciPostPhysCodeb.8,10.21468/SciPostPhysCodeb.8-r8.3} and \textsc{Delphes 3.4.1}~\cite{deFavereau:2013fsa}. 
It contains $10^6$ QCD dijet background events and $10^5$ signal events of a vector boson $W'$ decaying into two vector bosons $X$ and $Y$ with masses $m_{W'}=\SI{3.5}{\tera\electronvolt}$, $m_X=\SI{500}{\giga\electronvolt}$ and $m_Y=\SI{100}{\giga\electronvolt}$. Both $X$ and $Y$ decay hadronically into light quarks. The LHCO  R\&D data set contains two different signal models, with $X$ and $Y$ particles decaying either into two light quarks each (called LHCO 2-prong in the following) or into three light quarks each (LHCO 3-prong). 

Additionally, we use the \num{612858} QCD dijet background events in the SR from \cite{extraLHCOdataset} for testing and the construction of a background template for the IAD, as was done in Ref.~\cite{Hallin:2021wme}.

\subsection{Additional signal models}
\label{sec:additional_signal_models}

\begin{table}[t]
    \centering
    \def\arraystretch{1.5}
    \begin{tabular}{|c|c|c|c|}
         \hline
         Name & Prongs & Masses  & $\bar{S}/N_\text{sig}$ \\
         \hline
         \multirow{2}{*}{LHCO 2-prong} & \multirow{2}{*}{$2 + 2$} & $m_Y=\SI{100}{\giga\electronvolt}$, & \multirow{2}{*}{$75.30\%$} \\
         & & $m_X=\SI{500}{\giga\electronvolt}$ & \\\hline
         \multirow{2}{*}{LHCO 3-prong} & \multirow{2}{*}{$3 + 3$} & $m_Y=\SI{100}{\giga\electronvolt}$, & \multirow{2}{*}{$79.79\%$} \\ & & $m_X=\SI{500}{\giga\electronvolt}$ &  \\\hline
         \multirow{2}{*}{$X \to Y  Y^\prime \to 4 q$} & \multirow{2}{*}{$2 + 2$} & $m_Y=\SI{100}{\giga\electronvolt}$, & \multirow{2}{*}{$67.39\%$} \\ 
         & & $m_{Y^\prime}=\SI{500}{\giga\electronvolt}$ &  \\\hline
         $W_{KK} \to W  R \to 3W$ & $2 + 4$ & $m_R = \SI{500}{\giga\electronvolt}$ & $82.24\%$\\\hline
         $Z^\prime \to T^\prime T^\prime \to t Z t Z$ & $5 + 5$ & $m_{T^\prime}=\SI{400}{\giga\electronvolt}$ & $79.29\%$\\\hline
         $G_{KK} \to H H \to 4t$ & $6 + 6$ & $m_H=\SI{400}{\giga\electronvolt}$ & $74.58\%$\\
         \hline
    \end{tabular}
    \caption{Resonant signal models considered in this analysis and their main characteristics, including the overall prong structure (Prongs), the mass of the intermediate BSM particles (Masses), and the expected fraction of signal events in the signal region $\bar{S}/N_\text{sig}$, where $\bar{S}$ is the expectation value for the number of signal events in the signal region. A detailed description of the 
    models is given in Section~\ref{sec:Data}.}
    \label{tab:signals}
\end{table}

The LHCO R\&D data set is well studied, but limited in its signal variety. Following Ref.~\cite{CMS:2024nsz}, we expand this benchmark to more complex signal jet substructures by simulating four additional resonance models.
In contrast to \cite{CMS:2024nsz}, we only include hadronic decay channels.
An overview of the different signal models is given in Table~\ref{tab:signals}.

Each of the signal models follows the same basic topology as the LHCO R\&D benchmark, $A\to BC$, where a heavy resonance $A$ decays into two intermediate resonances $B$ and $C$. For simplicity, the masses of all the heavy resonances $A$ are set to $m_A=\SI{3.5}{\tera\electronvolt}$. 

\begin{enumerate}
\item The first signal $X \to Y  Y^\prime \to 4 q$ with $m_Y=\SI{100}{\giga\electronvolt}$ and $m_{Y^\prime}=\SI{500}{\giga\electronvolt}$ has a $2+2$ prong topology like the LHCO 2-prong signal. However, unlike in the LHCO signal where all resonantly produced particles are vector bosons, the $Y^\prime$ particle is a scalar in this model. $Y^\prime$ decays into a pair of bottom quarks while $Y$ can also decay into pairs of light quarks.

\item The second signal $W_{KK} \to W  R \to 3W$ with $m_R=\SI{500}{\giga\electronvolt}$ consists of a heavy Kaluza-Klein vector boson $W_{KK}$ decaying into a $W$ boson and a scalar radion $R$~\cite{Agashe:2016rle,Agashe:2017wss}.
The radion decays into two $W$ bosons. 
We analyze the fully hadronic channel, where all $W$ bosons decay into two light quarks each, resulting in a $2+4$ prong structure.

\item The third signal $Z^\prime \to T^\prime T^\prime \to t Z t Z$ with $m_{T^\prime}=\SI{400}{\giga\electronvolt}$ consists of a $Z^\prime$ vector boson that decays into two vector-like quarks $T^\prime$~\cite{Okada:2012gy,Buchkremer:2013bha}. 
The $T^\prime$ particles decay into a top quark and a $Z$ boson. 
Again we only consider the fully hadronic channel, where all the intermediate vector bosons decay into light quarks, resulting in a $5+5$ prong topology.

\item The last signal $G_{KK} \to H H \to 4t$ with $m_H=\SI{400}{\giga\electronvolt}$ consists of a heavy spin-2 Randall-Sundrum graviton $G_{KK}$ that decays into two Higgs-like scalars $H$ \cite{Carvalho:2014lsg}.
The scalars decay into two top quarks, resulting in a $6+6$ prong structure in the fully hadronic channel which is considered here.
\end{enumerate}

We use \madgraph\textsc{3.6.2}~\cite{Alwall:2014hca} at leading order to simulate the hard process, including the decay of all intermediate resonances so that the final state consists of quarks only. The width of the BSM resonances is chosen such that they are effectively treated in the narrow width approximation. The resulting parton-level events are passed to \textsc{Pythia 8.313} for parton showering and hadronization, followed by \textsc{Delphes 3.5.0} for detector simulation, using the same detector card as in the LHCO R\&D setup.
To ensure compatibility with the LHCO configuration, we use the default \textsc{Pythia} tune but switch off multi-parton interactions. We provide the simulated events and simulation cards for the different signal models in \cite{signal-zenodo}.

\subsection{Preprocessing}
For all data sets, we cluster the reconstructed particles using the anti-$k_T$ jet algorithm~\cite{Cacciari:2008gp}, implemented in \texttt{FastJet}~\cite{Cacciari:2011ma}, with a jet radius of $R = 1$. We select the two jets with the highest $p_T$, if the leading jet passes the threshold of $p_T > \SI{1.2}{\tera\electronvolt}$.
The jets are then reordered by mass such that jet 1 is the lighter of the two jets.

\subsection{Training and test sets}
\label{sec:TrainingTesting}

The construction of the training and test sets is standard and follows Refs.~\cite{Finke:2023ltw,Hallin:2021wme}. We define the signal region (SR) by $m_{JJ}\in(3.3,\ 3.7)$ $\SI{}{\tera\electronvolt}$. Hence, the resonances of all the tested signal models lie at the center of the SR. 

For an analysis with $N_\text{sig}$ signal events, we keep only the signal events in the SR and add background events such that $S+B=$ \num{120000}. 

To estimate statistical uncertainties, for each $N_\text{sig}$ value, we construct 10 independent SR samples by randomly selecting signal events from the simulated data and combining them with the LHCO background. For each training set, the number of signal events in the SR data for a given $N_\text{sig}$ is subject to statistical (Poisson) fluctuations.
We also initialize the classifier differently for each training set, resulting in different training and validation splits. To compare different feature sets at fixed $N_\text{sig}$, we use the same training sets for all feature sets.

For the IAD, the background template contains roughly \num{270000} background events in the SR taken from the additional background set provided by \cite{extraLHCOdataset}. For CWoLa hunting, the sidebands (SB) left and right of the SR are $\SI{0.2}{\tera\electronvolt}$ wide and contain $~\sim$\num{130000} events. 
To account for imbalanced training sets, we use class weights.
The independent test set contains $~\sim$ \num{340000} background events and $\sim$ \num{20000} signal events in the SR.

\section{Feature sets}
\label{sec:Feature_sets}

\begin{table*}[t]
    \centering
    \def\arraystretch{1.5}
    \begin{tabular}{|c|c|c|}
        \hline
        Name  & Size & Features\\
        \hline 
        \baseline & $4$ & $\{m_{J_1}, \ \Delta m_J, \ \tau^{(\beta=1)}_{21,J_1}, \ \tau^{(\beta=1)}_{21,J_2}\}$  \\
        \hline
        \multirow{2}{*}{\extended} & \multirow{2}{*}{$56$} & $\{m_{J_1}, \ \Delta m_J, \ \tau^{(\beta)}_{N,J_1}, \ \tau^{(\beta)}_{N,J_2}\}$ \\ 
        & & for $N\leq 9$ and $\beta \in \{0.5, 1, 2\}$ \\
        \hline
        \multirow{2}{*}{\efps} & \multirow{2}{*}{$980$} & $\{m_{J_1}, \ \Delta m_J, \ \text{EFP}_{i,J_1}^{(\beta=1)}, \ \text{EFP}_{i,J_2}^{(\beta=1)} \}$ \\
        & & for $i\in \{1, \dotsc, \ 489\}$ \\
        \hline
        \multirow{2}{*}{\kitchenSink} & \multirow{2}{*}{1034} & $\{m_{J_1}, \ \Delta m_J, \ \tau^{(\beta)}_{N,J_1}, \ \tau^{(\beta)}_{N,J_2}, \ \text{EFP}_{i,J_1}^{(\beta=1)}, \ \text{EFP}_{i,J_2}^{(\beta=1)} \}$ \\ 
        & & for $N\leq 9$, $\beta \in \{0.5, 1, 2\}$ and $i\in \{1, \dotsc, \ 489\}$ \\
        \hline
        \randomKitchenSink & \multirow{2}{*}{22} & $\{m_{J_1}, \ \Delta m_J\}$ plus 10 N-subjettiness and 10 EFP features \\ 
        (per GBDT model in ensemble) & & drawn at random from the \kitchenSink set  \\
        \hline
    \end{tabular}
    \caption{Feature sets considered for training.
    A detailed description of the feature sets is given in Section~\ref{sec:Feature_sets}.}
    \label{tab:feature-sets}
\end{table*}

All feature sets considered in this work include the mass of the lighter jet $m_{J_1}$ and the difference in the mass of the jets $\Delta m_J = m_{J_2} - m_{J_1}$. In addition, we use jet substructure observables based on N-subjettiness~\cite{Thaler:2010tr,Thaler:2011gf} and/or Energy Flow Polynomials (EFPs)~\cite{Komiske:2017aww} for the various feature sets.
We summarize the different sets in Table~\ref{tab:feature-sets} and outline their construction below.

\subsection{N-subjettiness-based features}

N-subjettiness observables~\cite{Thaler:2010tr,Thaler:2011gf} are widely used jet substructure variables that quantify to what extent a jet is consistent with being composed of N subjets. We compute N-subjettinesses using the \texttt{FastJet Contrib} implementation by the authors of~\cite{Thaler:2010tr,Thaler:2011gf} as in Ref.~\cite{Finke:2023ltw} and in the LHCO R\&D data set. 

We construct a \emph{\baseline} feature set following Refs.~\cite{Finke:2023ltw,Hallin:2021wme}. In addition to the mass features, this set consists of the N-subjettiness ratios $\tau_{21, J_h}^{(\beta=1)} = \tau_{2, J_h}^{(\beta=1)}/\tau_{1, J_h}^{(\beta=1)}$ for the two jets $h\in\{1,2\}$.

Ref.~\cite{Finke:2023ltw} introduced a larger N-subjettiness feature set using additional N-subjettinesses with $N\le 9$  and angular weighting factors $\beta\in\{0.5, 1, 2\}$.
This feature set has been shown to perform very well for the LHCO 2-prong and 3-prong signals. We will refer to this feature set as the \emph{\extended} set.

\subsection{Energy Flow Polynomials}

Energy Flow Polynomials (EFPs)~\cite{Komiske:2017aww} form a linear overcomplete infinite basis of all infrared- and collinear-safe observables. They allow for a flexible and expressive representation of jet substructure. We compute EFPs using the \texttt{energyflow} package~\cite{Komiske:2017aww,Komiske:2018cqr,Komiske:2019fks,Komiske:2019jim,Komiske:2019asc,Andreassen:2019cjw,Komiske:2020qhg}, employing the hadronic measure with energy fraction $z_i = p_{T, i}/p_{T, J}$, where $p_{T, i}$ and $p_{T, J}$ are the transverse momentum of the jet constituent $i$ and the whole jet, respectively. As for the N-subjettinesses, the angular distance $\theta_{ij}^{(\beta)} = [\Delta (y_{ij})^2 + \Delta (\phi_{ij})^2]^{\beta/2}$ between two jet constituents is defined based on rapidity ($y$) and azimuthal angle ($\phi$).

Different EFPs are characterized by a number of vertices and edges forming a loopless multigraph, where for prime EFPs, the corresponding graph is connected.  
Due to computational constraints, we truncate the infinite EFP basis to a set of prime EFPs with up to $d\leq7$ edges. Truncating the EFP basis this way results in \num{490} EFPs per jet, including EFPs that probe up to a prong structure of $4$.
For all of these EFPs, we choose a weighting factor of $\beta = 1$.
From the set, we exclude the trivial single-vertex EFPs, and we add the two mass features $m_{J_1}$ and $\Delta m_J$. Hence, the \emph{\efps} set includes \num{980} features.

\subsection{Kitchen Sink and random subsets}
\label{sec:KitchenSink}
We propose the ``kitchen sink'' approach, simply using everything at once by combining the mass features, the \extended set and the \efps set to gain more model independence. We call this composite of \num{1036} features the \emph{\kitchenSink} set.

The performance of GBDT models scales well with the feature set size, but its training time does increase linearly. 
To mitigate this, we leverage the ensemble structure of our classifier by training each GBDT on a random subset of features from a larger feature set.
Such a random selection of features for an ensemble of decision trees is well known as the random subspace method~\cite{709601} or attribute bagging~\cite{BRYLL20031291}. 
As a proof of concept for weakly supervised anomaly detection, we choose to randomly select \num{10} N-subjettinesses and \num{10} EFPs from the \kitchenSink set, while keeping the mass features fixed. 
Hence, each GBDT model is trained on a newly drawn subset of \num{22} features.
The random selection is done at runtime and therefore depends on the initialization of the classifier.
We refer to this configuration as the \emph{\randomKitchenSink} set. 

\section{Results}
\label{sec:Results}

For the different feature sets and signal models, we present the $\max(\text{SIC})$ as a function of $N_\text{sig}$ in Figures~\ref{fig:IADmaxSIC} and~\ref{fig:CWoLaResults}. We show the median and $68\%$ confidence interval by linearly interpolating the results for the 10 different training samples. We also show the expected initial significance $\bar{S}/\sqrt{B}$ for a given signal injection, where $\bar{S}$ is the expectation value for the number of signal events in the SR. As shown in Table~\ref{tab:signals}, $\bar{S}/ N_\text{sig}$ and hence the initial significance differ for the various signal models. 

We first discuss the performance of the feature sets on the LHCO signals for the IAD in Section~\ref{sec:ResultsIAD} and investigate the more complex signal models in Section~\ref{sec:IADBeyondLHCO}. We show that our findings are not specific to the IAD but also hold for CWoLa hunting in Section~\ref{sec:ResultsCWoLa}. Finally, we investigate the initial significance needed for $3\sigma$ evidence or a $5\sigma$ discovery in Section~\ref{sec:discoveries}.  

\subsection{IAD for the LHCO signals}
\label{sec:ResultsIAD}
The results for the LHCO 2-prong signal (Figure~\ref{fig:IADmaxSIC} upper left) can be directly compared to the results in Ref.~\cite{Finke:2023ltw}. We observe a slightly lower median performance for the \baseline feature set because the fixed choice of signal events used in Ref.~\cite{Finke:2023ltw} were slightly favorable for the performance. We confirm the findings in Ref.~\cite{Finke:2023ltw} that the \extended set is a much better feature set for identifying the anomalous LHCO 2-prong events. 

The \efps set even outperforms the \extended set for small $N_\text{sig}$, in particular the performance breaks down at a considerably smaller signal injection. Despite the large dimension of the \efps set, the decision trees identify the signal even for very small $N_\text{sig}$.
While Ref.~\cite{Finke:2023ltw} showed that also GBDT models eventually break down if too many Gaussian noise features are added, we find that many highly-correlated features with redundant 
information are not harmful.  

Combining the \extended and \efps feature sets, the resulting feature set performs as well as the \efps set. Features probing a 2-prong structure with a high number of edges for EFPs or a high $\beta$ value for N-subjettinesses are the main drivers of the performance in the various feature sets (see Appendix~\ref{appendixA}).

As discussed in Section~\ref{sec:KitchenSink}, the classifier setup allows us to randomly select a subset of features for each decision tree in the ensemble of 50 GBDTs. 
This approach decreases total run time by a factor of $\sim20$ and pure training time by a factor of $\sim50$ with respect to the \kitchenSink feature set. For the LHCO signals, it even  improves overall performance. While the GBDT classifiers are robust to many redundant features, in this case they nevertheless slightly benefit if the most useful features do not have to be identified in a large number of uninformative or redundant features.
We observe similar results when using a random selection from the \efps set, indicating that the \efps set profits most from the reduction to smaller subsets.

In general we find the same behavior for the LHCO 2-prong and 3-prong signals.
While the LHCO 3-prong signal tends to be harder to find, the overall ordering of feature sets is the same as for the 2-prong case. 

\subsection{IAD for other signal models}
\label{sec:IADBeyondLHCO}
\begin{figure*}[p]
    \centering
    \includegraphics[width=0.49\linewidth]{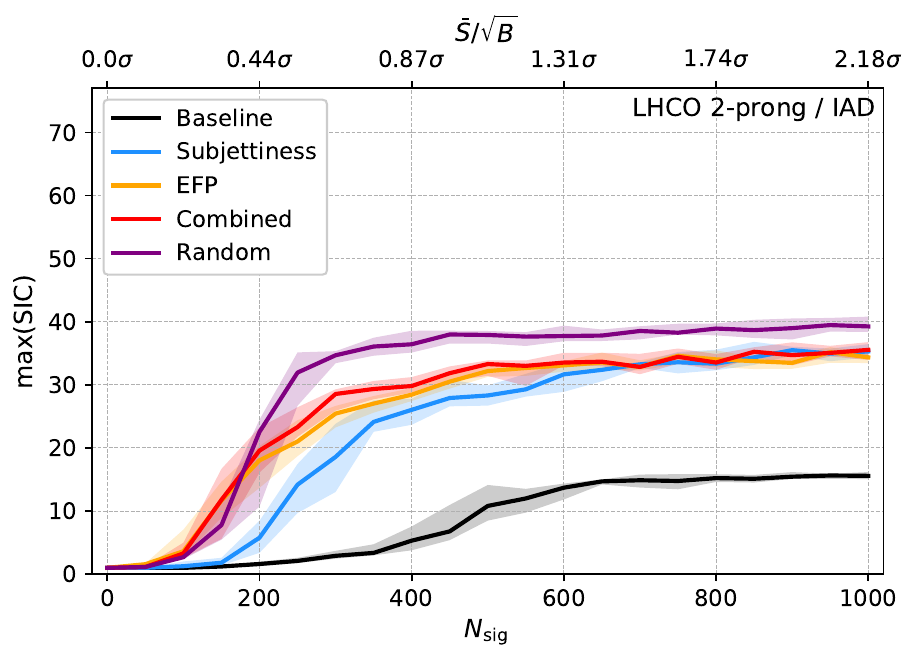}
    \includegraphics[width=0.49\linewidth]{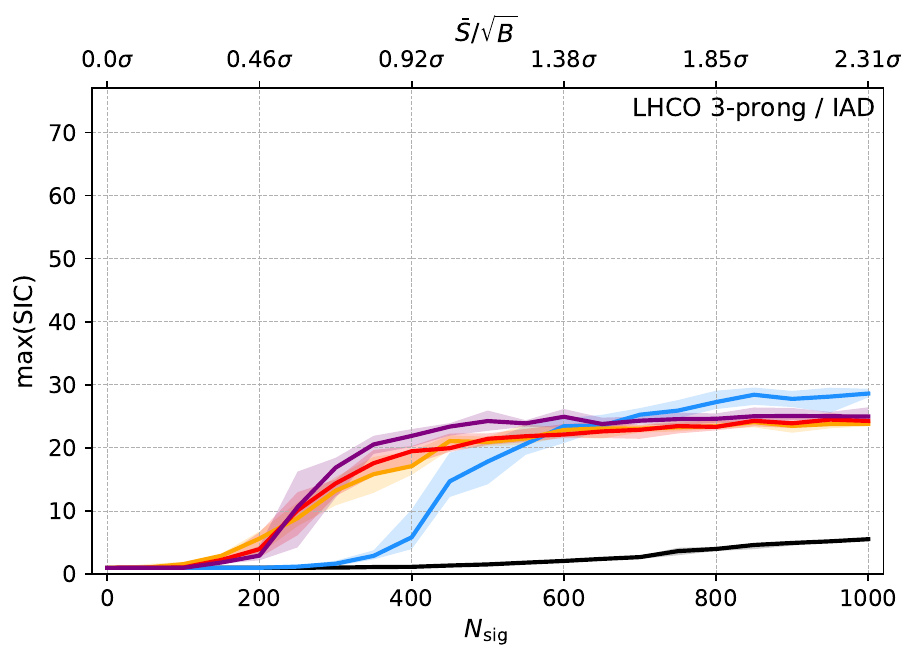}
    \includegraphics[width=0.49\linewidth]{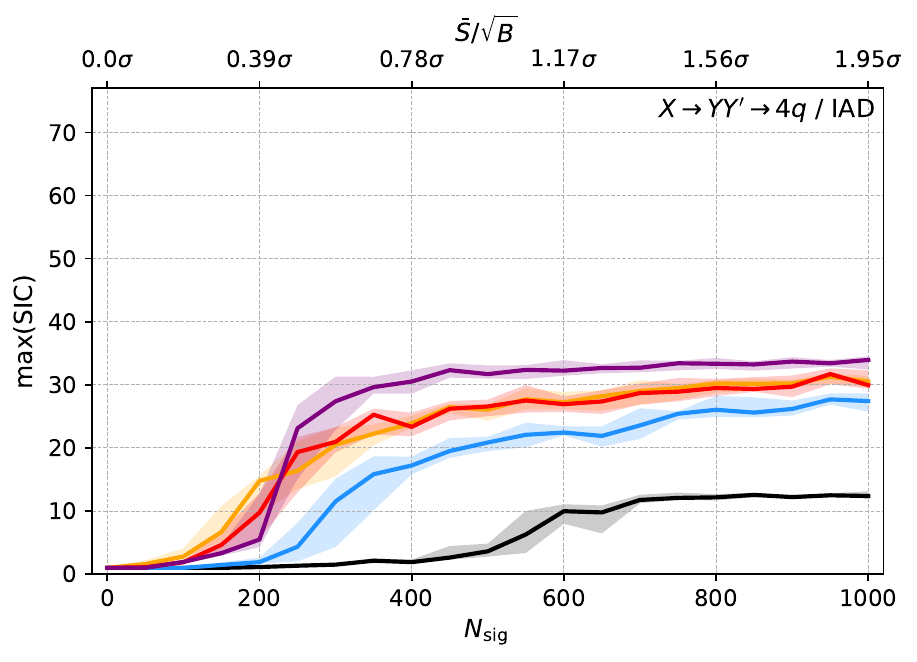}
    \includegraphics[width=0.49\linewidth]{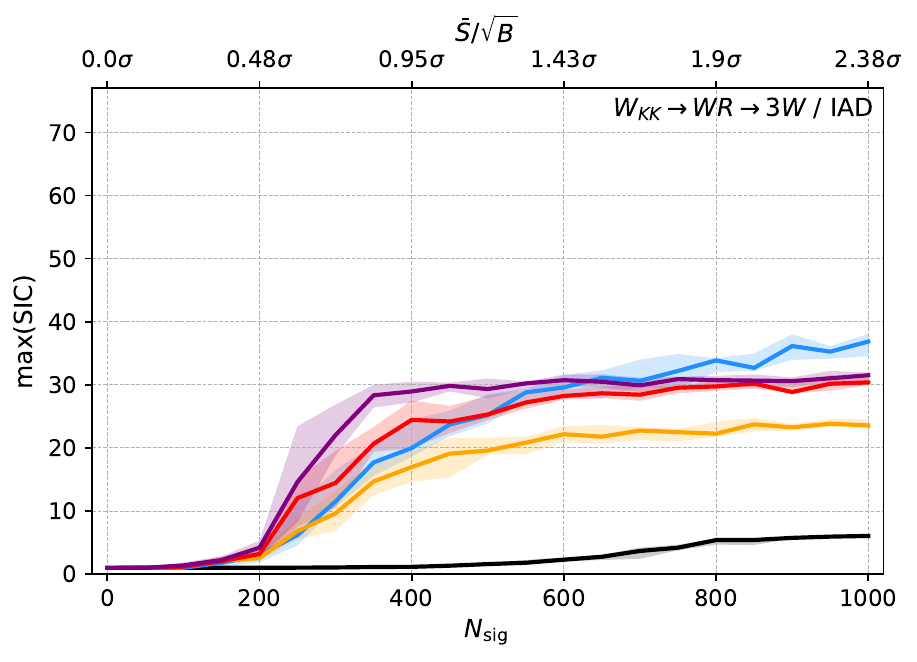}
    \includegraphics[width=0.49\linewidth]{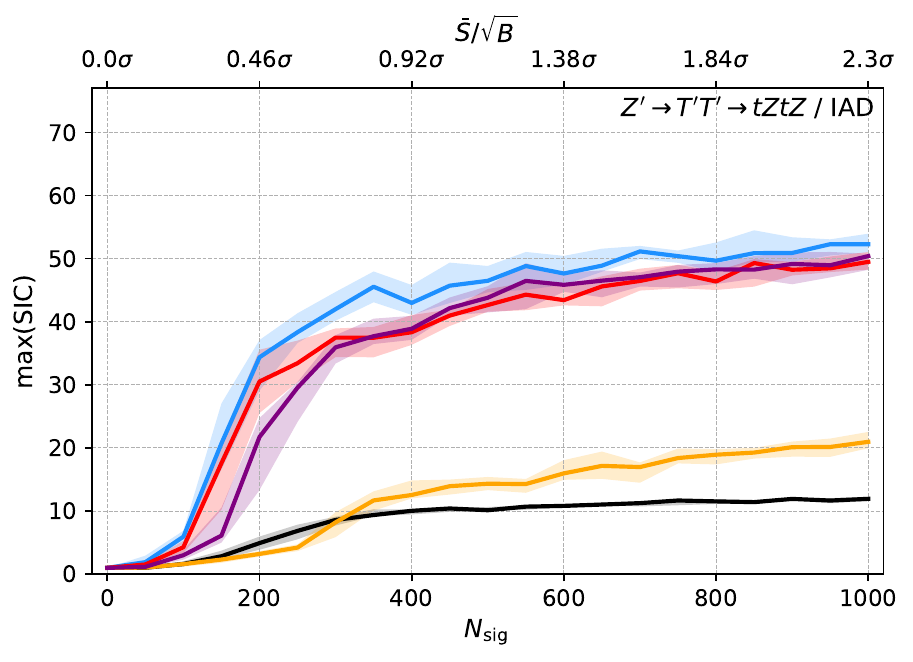}
    \includegraphics[width=0.49\linewidth]{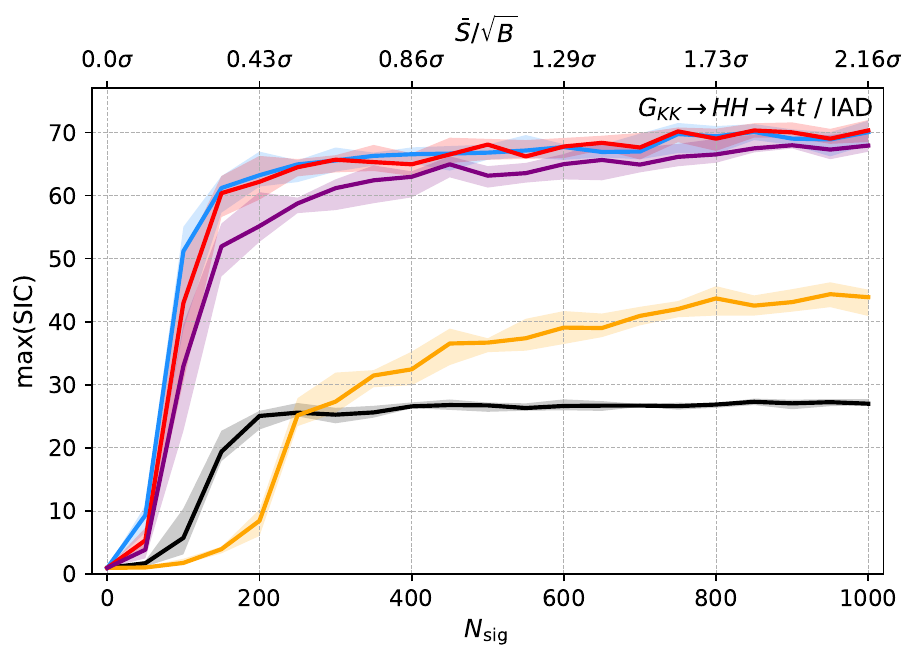}
    \caption{Maximum significance improvement characteristic (SIC), eq.~(\ref{eq:SIC}), of the different feature sets from Table~\ref{tab:feature-sets} as a function of the signal injection $N_\text{sig}$  for the various signal models of Table~\ref{tab:signals}, in the idealized anomaly detector (IAD) setup. 
    The corresponding expected initial significance $\bar{S}/\sqrt{B}$ is indicated on the upper horizontal axis.
    }
    \label{fig:IADmaxSIC}
\end{figure*}

As shown in Figure~\ref{fig:IADmaxSIC}, the \efps set performs very well on some of the signal models but not on all. This behavior can be understood from the different substructure patterns of the signal models.  EFPs probe multi-particle correlations with increasing sensitivity to large-angle radiation as the number of edges grows, making them particularly well suited for identifying well-separated prong structures such as in the LHCO signals. In contrast, for more complex signals with many decay products per jet, the radiation pattern becomes more isotropic in the rest frame of the intermediate resonance and less well described by a small number of dominant angular correlations.  In these cases, observables such as N-subjettiness, which effectively distinguish between one-prong and multi-prong or more diffuse radiation patterns, can provide a more robust characterization of the jet substructure.

If the \efps set is outperformed by the \extended set, the \kitchenSink feature set still achieves competitive performance and can be considered the most robust among all feature sets. Hence, we demonstrate the strength of the ``kitchen sink'' approach.

While the prong structure and the resonance masses of $X \to Y Y^\prime \to 4q$ are identical to the LHCO 2-prong signal, the spin-structure is different. We observe a decrease in overall performance for all feature sets. However, the overall ordering of feature sets is the same as observed for the LHCO signals.

In the case of the $W_{KK} \to WR \to 3W$ signal (see Figure~\ref{fig:IADmaxSIC} center right), the \randomKitchenSink feature set performs best. Here, the GBDT classifier indeed has some trouble identifying the most relevant features in the high dimensional sets. A random selection of only EFP features still outperforms the \kitchenSink set.

For the signal models $Z^\prime \to T^\prime T^\prime \to t Z t Z$ and $G_{KK}\to HH\to 4t$ with many prongs in each jet, the \extended set significantly outperforms the \efps set. For $G_{KK}\to HH\to 4t$, the $2$-subjettiness mainly drives the performance, which is consistent with the more isotropic radiation pattern of this signal (see Appendix~\ref{appendixA}). Nevertheless, the combined set of EFPs and subjettinesses performs similarly to the \extended set, showing that a combination of feature sets leads to optimal model agnostic performance. 

In the high signal injection regime, the IAD should approach the supervised classifier, and in this limit the feature set with the most information should perform best. Although the \kitchenSink set is still outperformed for some signals at $N_\text{sig}=1000$, we observe that it indeed performs best in the supervised limit. 

\subsection{Generalization to CWoLa hunting}
\label{sec:ResultsCWoLa}
\begin{figure*}[p]
    \centering
    \includegraphics[width=0.49\linewidth]{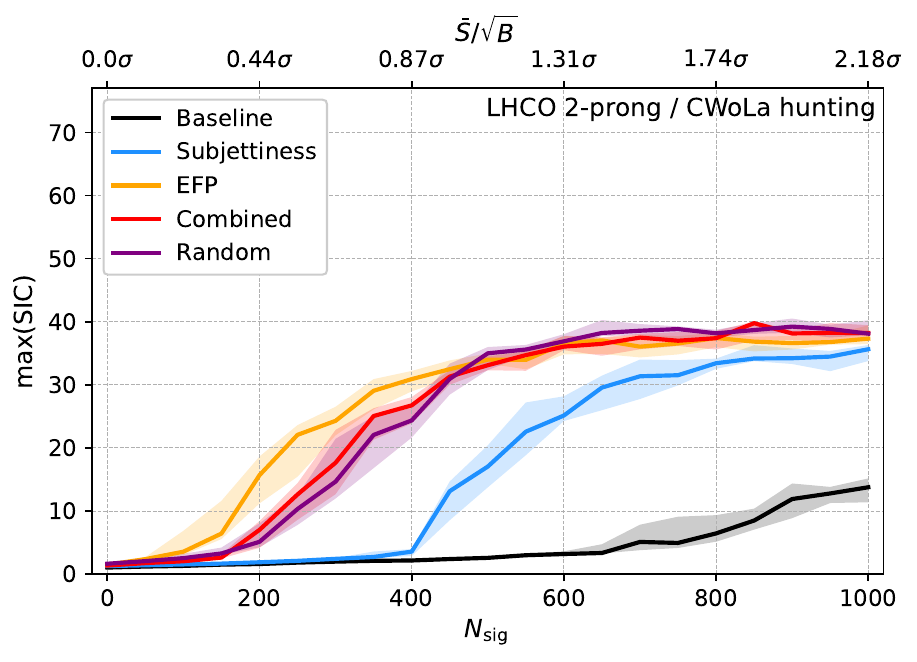}
    \includegraphics[width=0.49\linewidth]{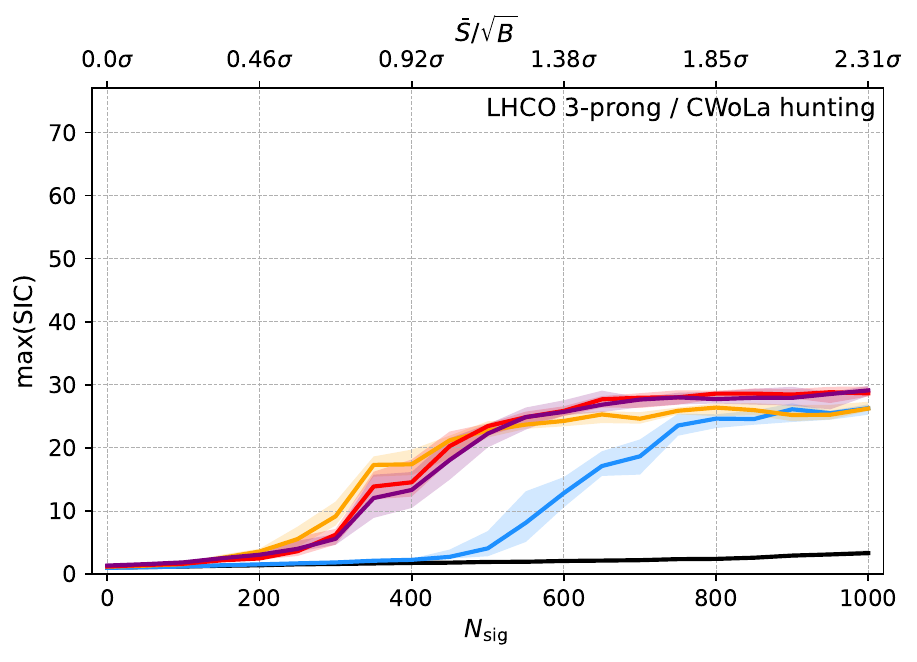}
    \includegraphics[width=0.49\linewidth]{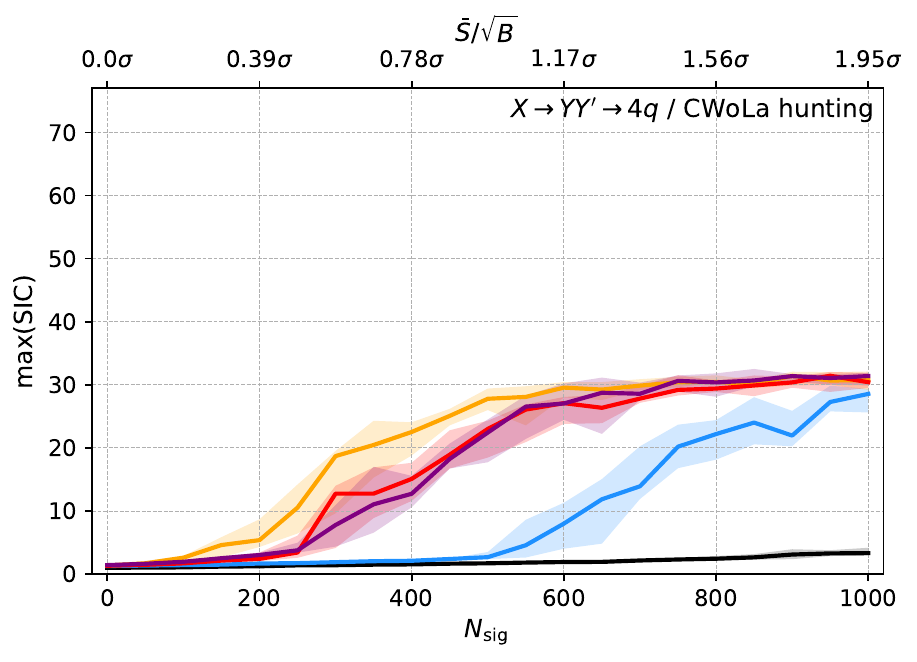}
    \includegraphics[width=0.49\linewidth]{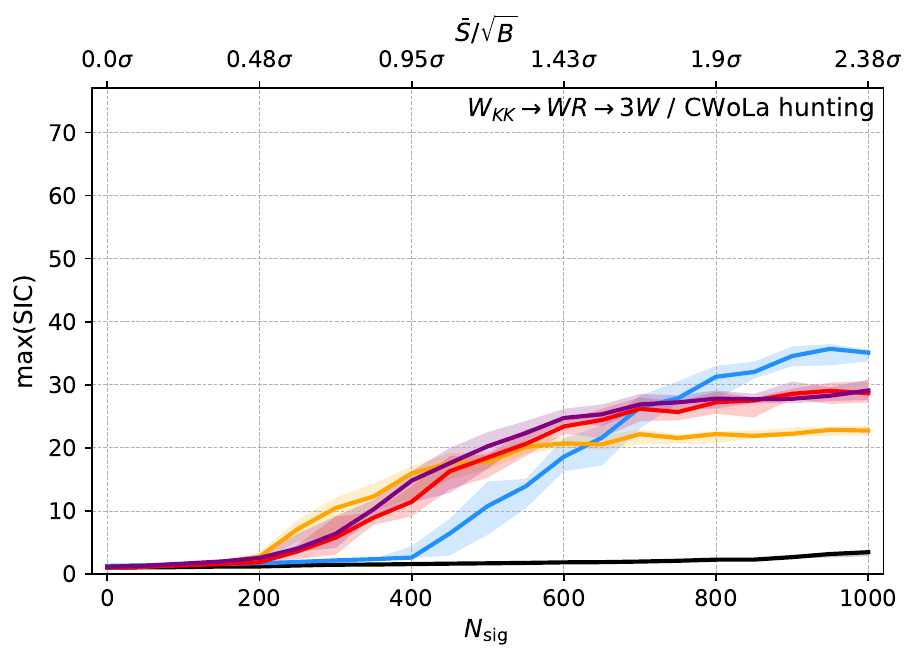}
    \includegraphics[width=0.49\linewidth]{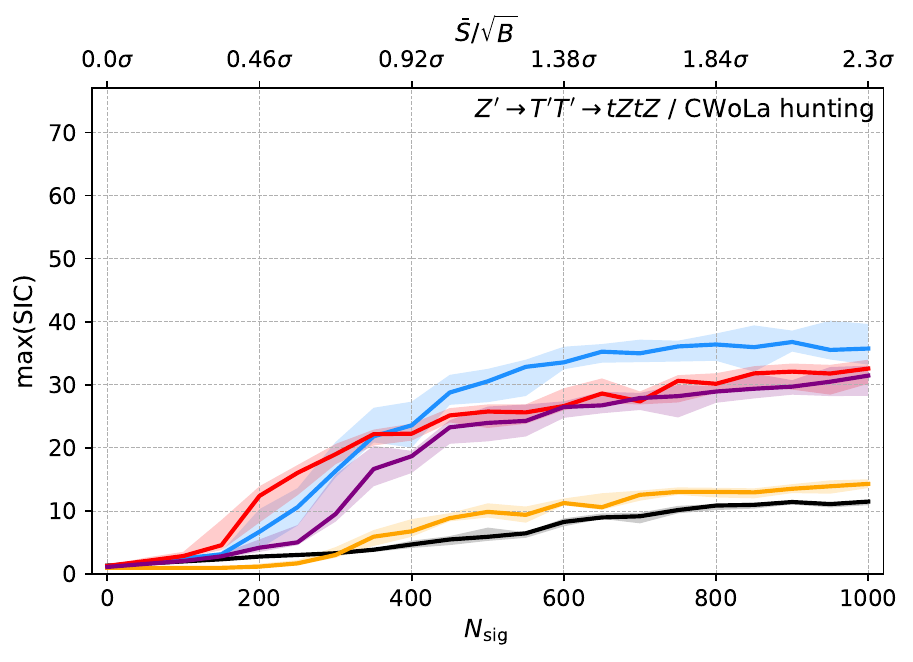}
    \includegraphics[width=0.49\linewidth]{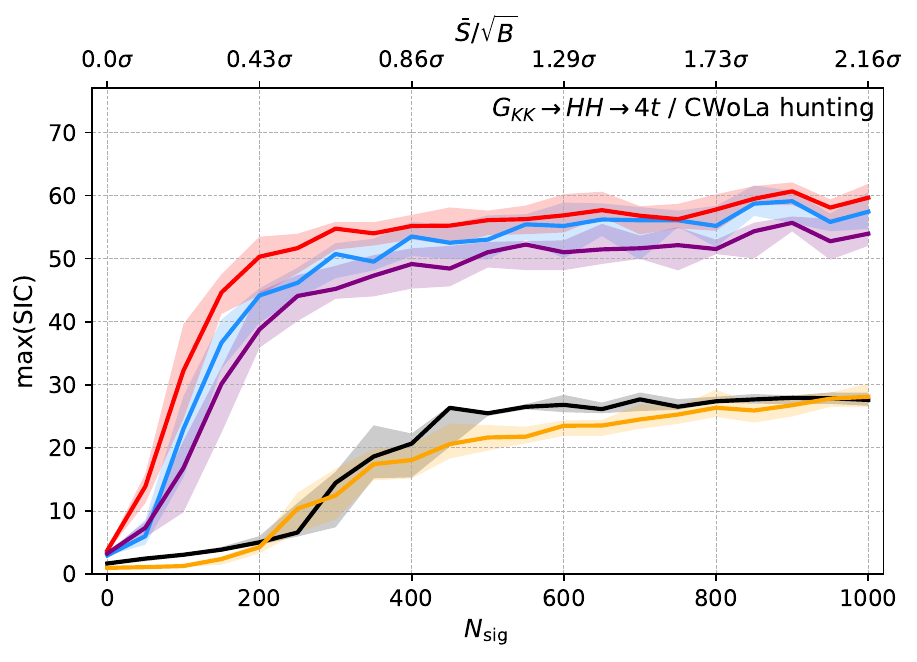}
    \caption{The same as for Figure~\ref{fig:IADmaxSIC} but for the CWoLa hunting setup.}
    \label{fig:CWoLaResults}
\end{figure*}

For CWoLa hunting (see Figure~\ref{fig:CWoLaResults}), the \kitchenSink set is again the most model agnostic and best performing feature set. 

As expected, overall performance decreases for CWoLa hunting in comparison to the IAD for all feature sets due to the imperfect BT. The performance drop for the \extended set and the \kitchenSink set is more severe than for the \efps set. While the \efps set performs best for some signal models, it shows significantly reduced performance for others and is therefore not the most model-agnostic choice.

The \randomKitchenSink set performs only slightly worse than the \kitchenSink set in most cases, resulting in a similar discovery potential while using  significantly less run time. 

In the case of the $G_{KK} \to HH \to 4t$ signal, the $\max(\text{SIC})$ is greater than 1 even for zero signal injection because the small differences between SB and SR data by chance allow the classifier to identify signal events. For the IAD, this behavior is of course not observed.

\subsection{Discoveries with the IAD and CWoLa hunting}
\label{sec:discoveries}
\begin{figure*}[p]
    \centering
    \includegraphics[width=0.965\linewidth]{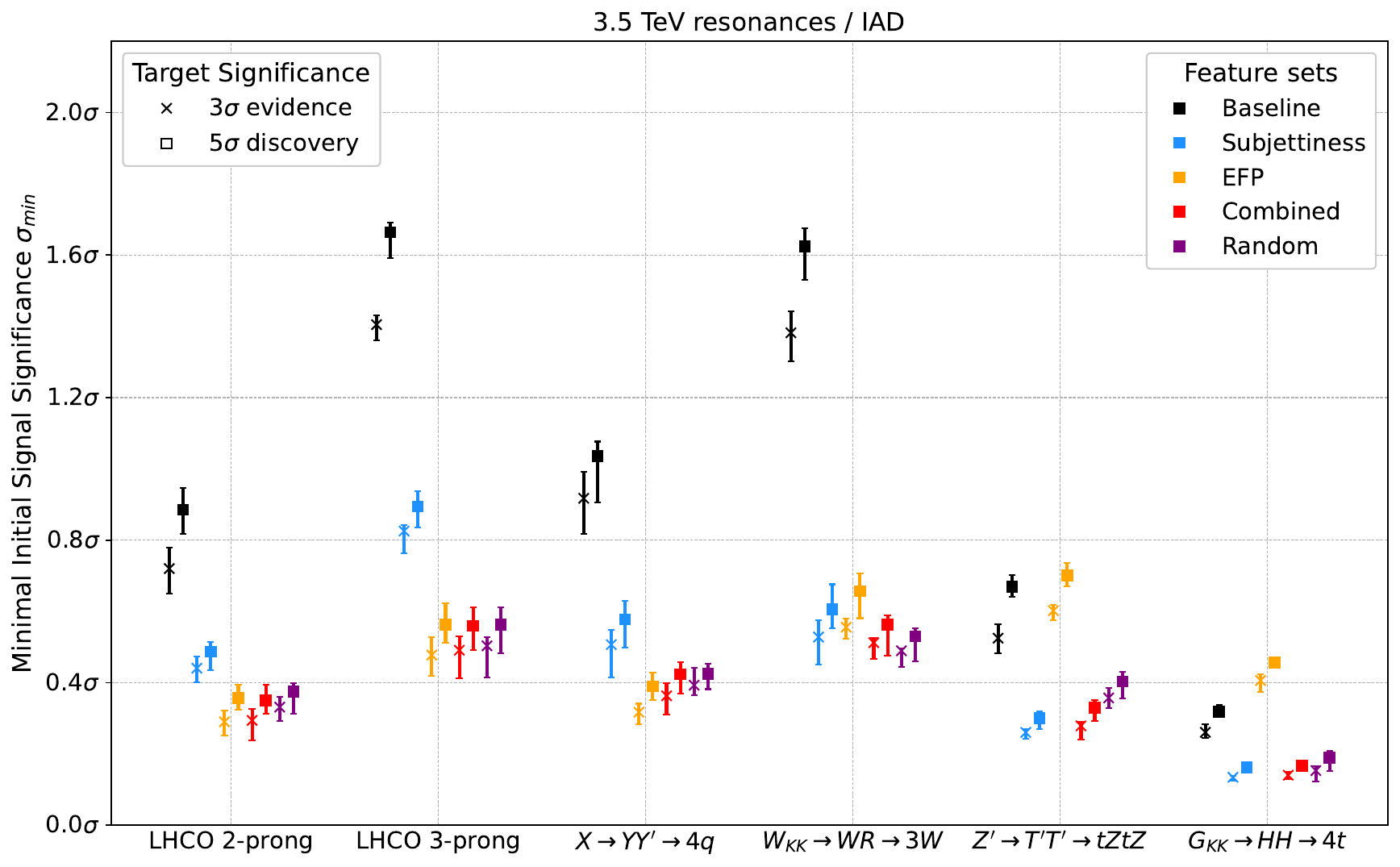}\\[5mm]
    \includegraphics[width=0.965\linewidth]{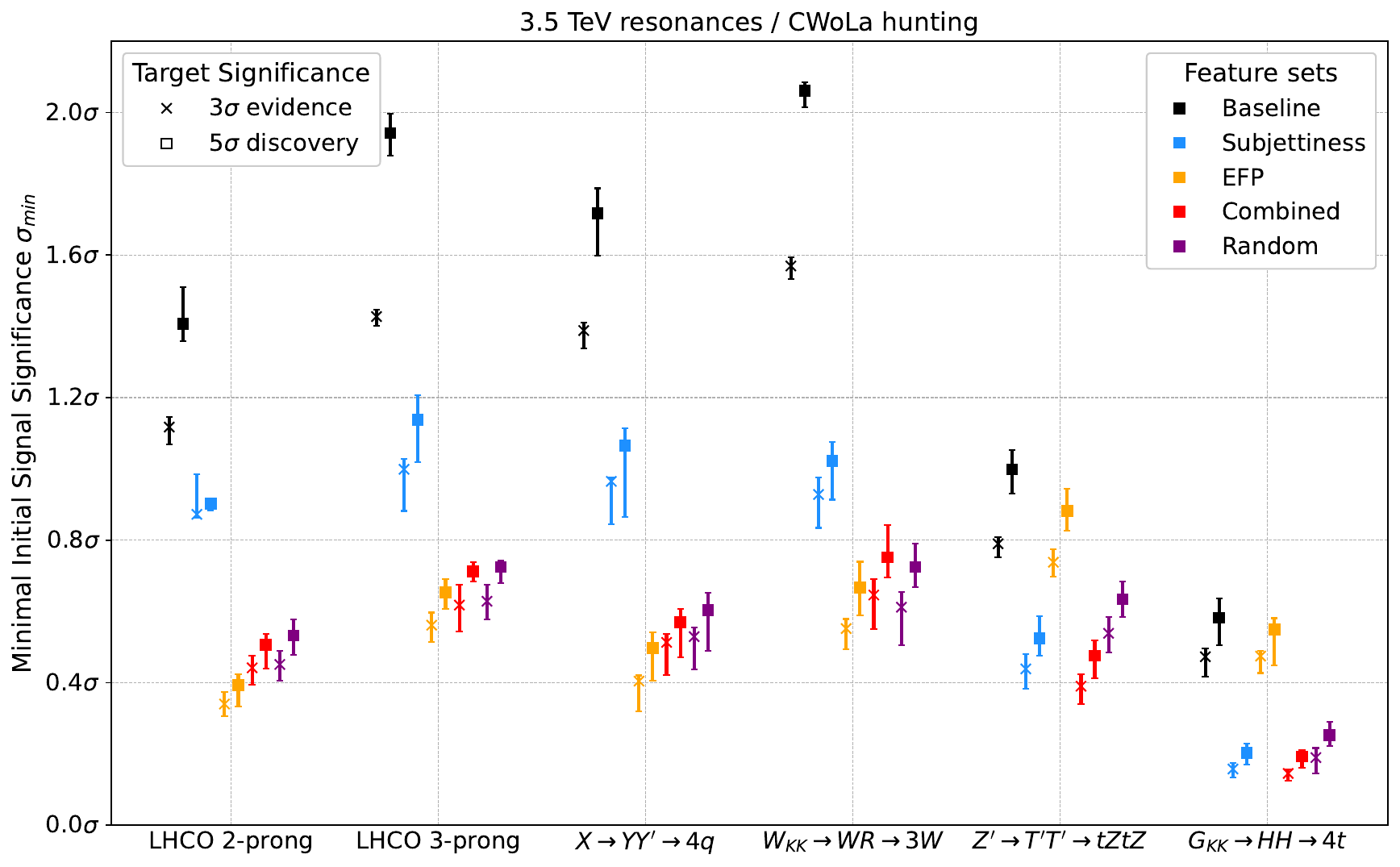}
    \caption{Minimal initial signal significance $\sigma_{min}$ (eq.~\ref{eq:minIS}) for $3\sigma$ evidence or a $5\sigma$ discovery for the different feature sets (see Table~\ref{tab:feature-sets}), across all investigated signal models (see Table~\ref{tab:signals}).
    The top plot shows the results for the IAD setup and bottom plot for the CWoLa hunting setup.
    }
    \label{fig:summary}
\end{figure*}

To summarize, we translate our findings into the discovery potential of the different feature sets. As discussed in Section~\ref{sec:Metrics}, for each classifier the significance is given by the maximum of the Asimov estimate (see eq.~(\ref{eq:AsimovStatistic})), without considering any systematic effects. We calculate the median of the significance for the 10 training sets and the corresponding $68\%$ confidence interval for each signal injection and feature set. 
The median and the error band for each feature set are then linearly interpolated to find the initial significance for which the Asimov estimate reaches 3 or 5$\sigma$. 
The results are shown in Figure~\ref{fig:summary} for the IAD (top) and the CWoLa hunting (bottom) setups.

These summary plots highlight that the \kitchenSink set performs best across all signal models. 
We quantify this statement by calculating the metrics discuses in Section \ref{sec:Metrics}. 
For a $5\sigma$ discovery, we present the average and maximum of the minimal initial significance eq.~(\ref{eq:minIS}) and regret eq.~(\ref{eq:regret}) for the different feature sets, over the investigated signal models in 
Table~\ref{tab:feature-set-comparison-IAD-5sigma}.
This highlights that the \kitchenSink feature set indeed  performs best across all signal models in the IAD and CWoLa hunting setting. 

\begin{table}[t]
    \centering
    \def\arraystretch{1.3}
    \begin{tabular}{|l|c|c|c|c|}
      \hline
      \multicolumn{5}{|c|}{IAD}\\
      \hline
      Feature set (f) & Avg $\sigma_{0,min}$ & Max $\sigma_{0,min}$& Avg $r_f$ & Max $r_f$\\
      \hline
      \baseline & \rankcell{5}{1.03} & \rankcell{5}{1.66}  & \rankcell{5}{2.57} & \rankcell{5}{3.07} \\
      \extended & \rankcell{3}{0.50} & \rankcell{4}{0.89}  & \rankcell{3}{1.27} & \rankcell{3}{1.60} \\
      \efps & \rankcell{4}{0.52} & \rankcell{3}{0.70} & \rankcell{4}{1.57} &   \rankcell{4}{2.82} \\
      \kitchenSink & \rankcell{1}{0.40} & \rankcell{1}{0.56}  & \rankcell{1}{1.05} & \rankcell{1}{1.10} \\
      \randomKitchenSink & \rankcell{2}{0.41} & \rankcell{1}{0.56} &   \rankcell{2}{1.11} & \rankcell{2}{1.34} \\
      \hline
      \multicolumn{5}{c}{}\\
      \hline
      \multicolumn{5}{|c|}{CWoLa hunting}\\
      \hline
      Feature set (f) & Avg $\sigma_{0,min}$ & Max $\sigma_{0,min}$ & Avg $r_f$ & Max $r_f$\\
      \hline
      \baseline & \rankcell{5}{1.45} & \rankcell{5}{2.06} & \rankcell{5}{3.04} & \rankcell{5}{3.58}\\
      \extended & \rankcell{4}{0.81} & \rankcell{4}{1.14} & \rankcell{4}{1.65} & \rankcell{3}{2.29} \\
      \efps & \rankcell{3}{0.61} & \rankcell{3}{0.88} & \rankcell{3}{1.45} & \rankcell{4}{2.85}\\
      \kitchenSink & \rankcell{1}{0.53} & \rankcell{2}{0.75} & \rankcell{1}{1.11} & \rankcell{1}{1.29} \\
      \randomKitchenSink & \rankcell{2}{0.58} & \rankcell{1}{0.73} & \rankcell{2}{1.23} & \rankcell{2}{1.35}\\
      \hline
    \end{tabular}
    \caption{Comparison of feature set (f) performance using various metrics in the IAD setup (top) and CWoLa hunting setup (bottom). Across the different signal models,
    we present the average and the maximum of the median minimal initial signal significance ($\sigma_{min}$ see eq.~(\ref{eq:minIS})) to reach a $5\sigma$ discovery and the average and maximum regret ($r_f$) as defined in eq.~(\ref{eq:regret}).
    The color represents the ranking for a given metric, from \textcolor{rank1}{dark green} (best) to \textcolor{rank5}{red} (worst).
    }
    \label{tab:feature-set-comparison-IAD-5sigma}
\end{table}

\section{Conclusion}
\label{sec:Conclusion}

We have demonstrated that weakly supervised resonant anomaly detection based on high-level observables can easily cope with $\mathcal{O}(1000)$ physically motivated features using boosted decision trees as the classifier architecture. This allows us to combine well-known feature sets based on $N$-subjettinesses with even larger feature sets consisting of energy flow polynomials. 

To investigate how model agnostic different feature sets are, we have applied our weakly supervised setup to several well motivated benchmark signals, where a BSM resonance decays via various hadronic decay channels into two fat jets. We have made these additional benchmark signal models publicly available on Zenodo~\cite{signal-zenodo}. We hope that the community will find these useful for method development.

Selecting specific feature sets for any given signal model allows for optimal performance but is of course not in accordance with the model agnostic paradigm. Instead, we suggest to use kitchen sink anomaly detection, i.e.\ to use all available features at once, to achieve state-of-the-art performance for all considered signal models. Since for BDTs computing time scales linearly with the number of features, to speed up anomaly detection one can use random subsets of features for the individual BDTs in our ensembling approach. The price for the speed-up in terms of loss of performance is generally small. 

Doing anomaly detection in the most model agnostic way is still a long-term goal. Here, we have put forward a practical alternative in the CWoLa hunting approach which can achieve state-of-the-art performance with currently available machine learning architectures for a broad variety of signal models down to small signal numbers.
The expansion of this approach to methods with interpolative constructions of the background template, such as CATHODE~\cite{Hallin:2021wme} or CURTAINs~\cite{Raine:2022hht}, poses a promising future challenge.

\section*{Acknowledgements}
MH is supported by the Deutsche Forschungsgemeinschaft (DFG, German Research Foundation) under grant 400140256 - GRK 2497: The physics of the heaviest particles at the Large Hadron Collider. 
The research of LL, MK and AM is supported by the DFG under grant 396021762 - TRR 257: Particle Physics Phenomenology after the Higgs Discovery. R.M. received support through Schmidt Sciences, LLC. GK and LM acknowledge support by the DFG under Germany’s Excellence Strategy 390833306 – EXC 2121: Quantum Universe. DS is supported by DOE grant DOE-SC0010008. RD acknowledges support from the Alexander
von Humboldt Foundation. Computations were performed with computing resources granted by RWTH Aachen University under project rwth0934.

\section*{Code}
The code for this paper can be found at \cite{Lang2026}. 
Data is provided in \cite{signal-zenodo} including the simulations cards and settings used.

\vspace{2mm}
\appendix
\section{Additional feature analysis}
\label{appendixA}

\begin{figure*}[t]
    \centering
    \includegraphics[width=0.49\linewidth]{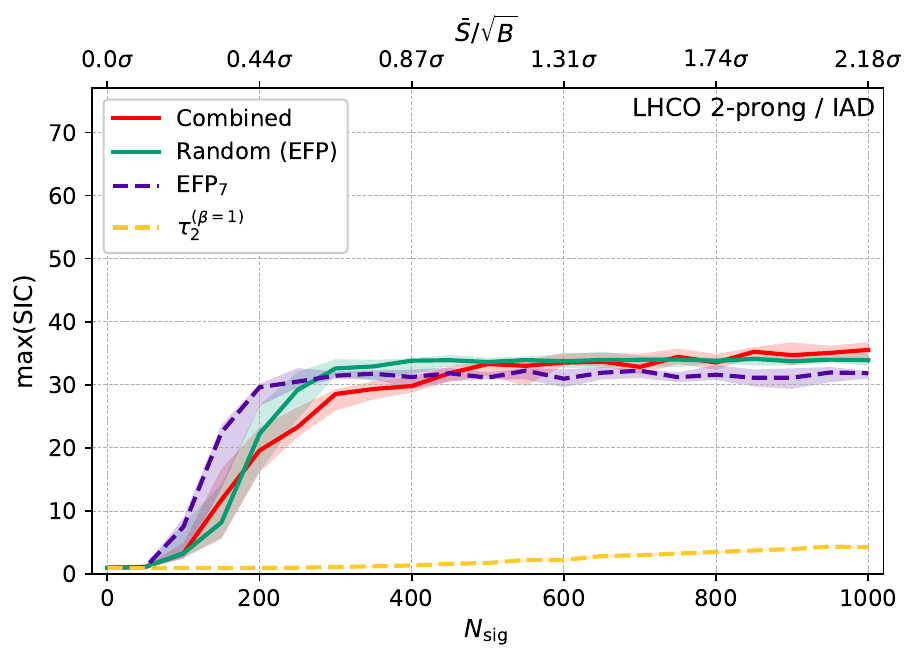}
    \includegraphics[width=0.49\linewidth]{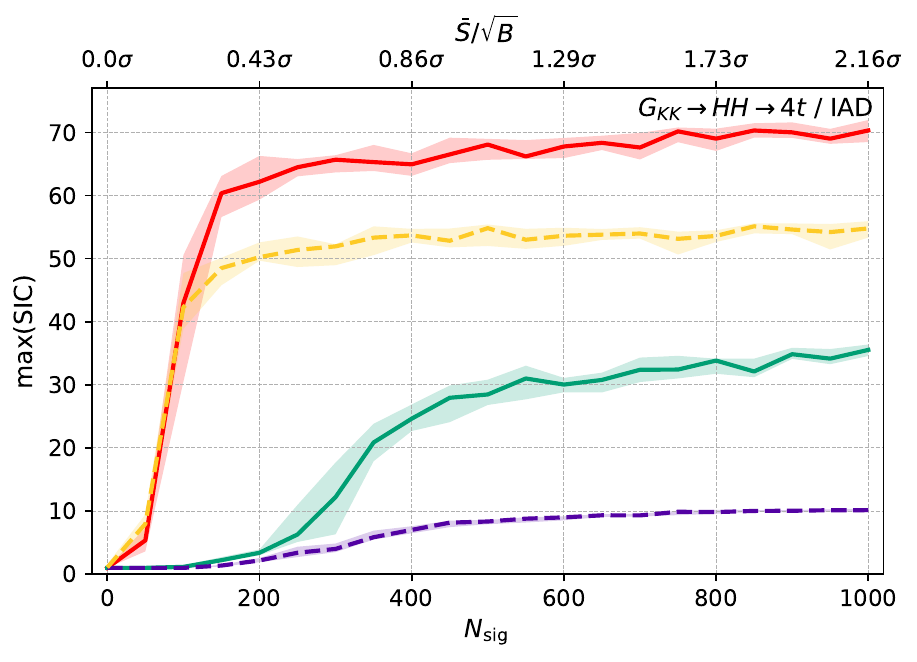}
    \caption{We present the performance of a selection of feature sets in the same setting as Figure~\ref{fig:IADmaxSIC} for the LHCO 2-prong and $G_{KK} \to HH \to 4t$ signal models. For reference we show the performance of the \kitchenSink set. The \randomEfps set is constructed similarly as the \randomKitchenSink set, with the \efps set as basis. The EFP$_7$ set and $\tau_{2}^{(\beta=1)}$ set contain the mass features $m_{J_1}$ and $\Delta m_{J}$ as well as, for each jet, the EFP$_7$ and $\tau_{2}^{(\beta=1)}$ features, respectively.}
    \label{fig:smalSets}
\end{figure*}

In this appendix, we present results for additional feature sets which perform particularly well for a given signal model but are not model agnostic. As before, all considered feature sets include the two mass features. In Figure~\ref{fig:smalSets}, we show exemplary results for the LHCO 2-prong and the $G_{KK} \to HH \to 4t$ signal models.

For the LHCO 2-prong signal, the best performance at small signal injections can be achieved by only adding one specific EFP per jet to the mass features. The EFP has 2 vertices and 7 edges, i.e.\ the angular distance enters with the largest power available in our truncated EFP set. Apparently, the LHCO 2-prong signal can be particularly well distinguished from the background by focusing on constituents with the largest angular separation. This is, of course, very specific to this signal model. For other signals like $G_{KK} \to HH \to 4t$, this EFP shows no performance at all.

The opposite behavior is observed when using only $\tau_2^{(\beta=1)}$ per jet in addition to the mass features. Although the $G_{KK} \to HH \to 4t$ signal consists of two jets with a 6-prong structure each, this minimal feature set performs very well. This can be understood from the kinematics of the signal. The Higgs-like scalar is produced only moderately above the $t\bar{t}$ threshold, such that the top quarks are only mildly boosted. Their hadronic decay products therefore populate the jet over a wide range of angles, resulting in a rather unstructured, quasi-spherical radiation pattern. In this case, the discrimination problem effectively reduces to separating one-prong QCD jets from more isotropic jets, which can already be captured by low-order subjettiness observables such as $\tau_2$. In contrast, the truncated EFP basis is less directly sensitive to this coarse structure.

In Figure~\ref{fig:smalSets}, we also show results for a random set for which 10 features are selected per GBDT only from the \efps set. While this set performs very well on the LHCO 2-prong signal, it does not for $G_{KK} \to HH \to 4t$. The same trend is observed for the other signal models: if the \efps set performs well, this is also the case for the corresponding random \efps set.

\bibliography{HEPML, other, models}
\bibliographystyle{apsrev4-1}

\end{document}